\documentclass[lettersize,journal]{IEEEtran}
\usepackage{amsmath,amsfonts}
\usepackage{algorithmicx}
\usepackage{algpseudocode}
\usepackage{algorithm}
\usepackage{array}
\usepackage[caption=false,font=normalsize,labelfont=sf,textfont=sf]{subfig}
\usepackage{textcomp}
\usepackage{stfloats}
\usepackage{multirow}
\usepackage{url}
\usepackage{verbatim}
\usepackage{makecell}
\usepackage{graphicx}
\usepackage{cite}
\usepackage{booktabs}
\usepackage[usenames,dvipsnames]{color}

\begin{document}

\title{SIGMA: Saliency-Guided Sparse Mask Attacks for Speech Emotion Recognition}

\author{Qiyang Sun, Yi Chang*, Zixing Zhang \IEEEmembership{Senior Member, IEEE}, and Björn W.\ Schuller \IEEEmembership{Fellow, IEEE}

\thanks{This research was supported by the National Natural Science Foundation of China under Grant No.\ 62571184.}
\thanks{Qiyang Sun and Yi Chang are with GLAM, Department of Computing, Imperial College London, UK. e-mail:q.sun23@imperial.ac.uk; yichang312@gmail.com.}
\thanks{Zixing Zhang is with the College of
Computer Science and Electronic Engineering, Hunan University, China; Zixing Zhang is also with the Shenzhen Research Institute, Hunan University, China. e-mail: zixingzhang@hnu.edu.cn.}
\thanks{Björn W.\ Schuller is with GLAM, Department of Computing, Imperial College London, UK; CHI -- Chair of Health Informatics, Technical University of Munich, Germany; relAI -- the Konrad Zuse School of Excellence in Reliable AI, Germany; MDSI -- Munich Data Science Institute, Germany; and MCML -- Munich Center for Machine Learning, Germany. e-mail: bjoern.schuller@imperial.ac.uk.}

\thanks{Corresponding author: Yi Chang.}
}

\markboth{Journal of \LaTeX\ Class Files,~Vol.~14, No.~8, August~2021}%
{Shell \MakeLowercase{\textit{et al.}}: A Sample Article Using IEEEtran.cls for IEEE Journals}


\maketitle

\begin{abstract}

Speech conveys rich emotional information. As Speech Emotion Recognition (SER) is usually deployed in privacy-sensitive and reliability-critical environments, adversarial attacks on SER have attracted increasing attention. Existing sparse attacks control the number of perturbed elements, yet,
they often lack explainability guidance and explicit measures of explanation consistency. A unified treatment of sparsity and magnitude constraints is also uncommon. In addition, transferability across attack families and target models remains limited. Hence, we propose a SalIency-Guided sparse Mask Attack (SIGMA). On self-supervised speech features, we use post-hoc explainable artificial intelligence (XAI) techniques to produce saliency maps and identify the scope of the mask, and then restrict magnitude-bounded updates to this mask. The mask is computed once and can be reused across models and different sparsity attacks to amortise cost.  We evaluate on 
the IEMOCAP and TESS datasets. Under matched budgets and across multiple sparse-attack settings, SIGMA maintains competitive attack success rates, navigating a conscious trade-off between attack efficacy and explanation consistency. SIGMA therefore provides an efficient and interpretable framework for analysing the vulnerability and explanation behaviour of SER models under structured perturbations.
\end{abstract}

\begin{IEEEkeywords}
Speech emotion recognition, adversarial attacks, explainable
artificial intelligence, sparsity,
transferability.
\end{IEEEkeywords}

\section{Introduction}
\IEEEPARstart{S}{peech} emotion recognition (SER) is the task of inferring a speaker’s affective state from acoustic and phonetic cues in the speech signal \cite{li2025gatedxlstm}. As a core component of affective computing, SER is valuable in privacy-sensitive and reliability-critical applications. It is widely used in mental-health screening, video-game interaction, and virtual agents services \cite{li2021make,matsouliadis2025speech,hu2022acoustically}. Its outputs can influence downstream decisions and user experience, making the reliability and explainability of SER systems important considerations. Looking towards Artificial General Intelligence (AGI) and Friendly AI, affect sensing and accurate empathy are regarded as key capabilities \cite{sun2024towards}. Robustness and reliability in SER are therefore essential.

Adversarial attacks expose the vulnerability of deep learning models to small perturbations and are now a standard tool for probing and improving robustness \cite{ren2020adversarial}. The area is most attractive in computer vision, with established methods and evaluation protocols \cite{8294186}. Subsequent studies show that small perturbations can also degrade speech systems, including automatic speech recognition \cite{zhang2022adversarial}, speaker recognition \cite{lan2022adversarial}, and SER \cite{ren2020generating}. In SER, subtle perturbations can alter affect predictions and threaten content moderation and interaction safety. In mental-health screening, they may lead to misclassification and inappropriate interventions. Strengthening the robustness of SER is therefore a practical priority. Yet, research on adversarial methods for SER remains limited. This does not mean that SER is hard to attack; rather, systematic benchmarks and methodologies are lacking on both attack and defense.

Adversarial attacks can be grouped by perturbation density into dense and sparse variants \cite{ghosh2022black}. Some attempts at sparsity attacks exist in audio tasks. Weighted-Sampling \cite{liu2020weighted} concentrates changes on a small number of frames or samples to reduce the noise rate and generation time. It relies on heuristic weights and denoising regularisers rather than on model attributions. Consequently, the modification locations lack a principled rule. As a result, perturbation locations often drift across models, and the setting targets attack success rates (ASR) rather than SER. The Audio Injection Attack \cite{liu2021audio} inserts very short perturbations into silence segments to gain temporal sparsity and stealth. Its support set is a “silence prior”, not evidence derived from the current input. For SER, which depends on cross-frame prosody and semantic cues, silence does not necessarily mark decision-critical regions, so effectiveness and interpretability are not guaranteed. Even within SER, STAA-Net \cite{chang2024staa} emphasises sparsity, transferability, and single-forward efficiency. It reduces run-time by training a generator to produce sparse perturbations. However, it is a specific attacker rather than a pluggable upper-layer constraint, and it is hard to reuse across different families of sparse attacks.

In white-box cross-model settings, mainstream iterative attacks must recompute gradients and re-optimise perturbations for each target model \cite{dong2018boosting}. The one-off cost is hard to amortise. In audio, recent studies on speech foundation models show that naively transferring an adversarial segment crafted on one model to another rarely succeeds, underscoring the difficulty of cross-model reuse and the overhead of repeated optimisation \cite{raina2024muting}. Other work reports that most adversarial examples are crafted around a single model, which yields weak cross-model transfer and requires extra mechanisms to boost transferability, creating a gap with practical needs \cite{jin2025boosting}. Thus, even under white-box transfer, the cost and stability of cross-model attacks remain unsatisfactory for real-world use.

Moreover, some studies combine explainable artificial intelligence (XAI) with adversarial attacks. The Focus-Shifting Attack \cite{huang2023focus} includes the saliency map in the loss. It first derives a manipulated mask from the original saliency, then optimises such that
the post-attack saliency matches this mask while constraining the change in logits. The goal is to mislead explanations, not to align perturbations with the model’s original evidence. The Maximal Jacobian-based Saliency Map Attack (JSMA) \cite{wiyatno2018maximal} builds a Jacobian saliency map from gradient sensitivity and iteratively selects a small set of coordinates. Its notion of saliency reflects local sensitivity to logits rather than post-hoc attribution, and it does not measure explanation consistency. The Saliency Attack \cite{dai2023saliency} operates in a black-box senario. It uses saliency object detection to localise visually salient regions and refines perturbations within them to improve imperceptibility. Its saliency is perceptual, not model attribution. Furthermore, most research is evaluated on image datasets. To the best of our knowledge, there is little work that treats explanation consistency as both a generation constraint and an evaluation dimension in audio or speech.

To address these gaps, we present \textbf{SIGMA}, a \textbf{S}al\textbf{I}ency-\textbf{G}uided sparse \textbf{M}ask \textbf{A}ttack framework for SER. SIGMA serves as a general and reusable saliency-guided constraint mechanism for existing sparse adversarial attacks. It uses post-hoc XAI techniques to select a top-k mask on audio features and restricts updates bounded in \(\ell_\infty\) to this mask, which controls sparsity and magnitude. We emphasise that our study is conducted in the latent feature space of self-supervised speech encoders. This setting serves as a controlled proxy for analysing model behaviour, rather than assuming direct access to internal representations in real-world deployments. In typical SER pipelines, SSL encoders act as fixed front-ends, and downstream classifiers operate on these features. Studying perturbations in this space allows us to isolate the role of salient feature components in model predictions and explanations. It has the following advantages:
\begin{enumerate}
  \item \textit{Pluggable.} SIGMA acts as an upper-layer constraint module and requires no change to the attack core. Experiments show effectiveness with different sparsity attacks.
  \item \textit{Transferable.} The saliency mask is generated once from the input and can be reused across models, which reduces transfer cost. Experiments show whitebox to transfer attacks across classifiers; at matched budgets the transfer ASR is not lower than the baseline and is higher in some cases.
  \item \textit{Explainable.} XAI guidance aligns perturbations with model evidence. Experiments show that compared with the baseline method, SIGMA can significantly improve the consistency of explanations.
\end{enumerate}

\section{Related Work}
\subsection{Speech Emotion Recognition (SER)}
In recent years, SER has evolved from hand-crafted or often specifically Mel-spectrogram features with traditional classifiers to deep learning (DL) methods. Early methods used engineered acoustic features as input and then applied classifiers for emotion recognition \cite{schuller2003hidden}. With the rise of DL, research in audio has shifted towards foundation models \cite{akman2025audio}. These models are pre-trained on large amounts of unlabelled speech. They are then fine-tuned for downstream emotion recognition or used as frozen feature extractors, which reduces labelling needs and improves robustness. Representative foundation models in audio tasks include wav2vec 2.0 \cite{baevski2020wav2vec},  WavLM \cite{wang2023speech},  and HuBERT \cite{yang2021superb}, also with some variants on SER,  including Emotion2Vec \cite{ma2023emotion2vec} and ExHuBERT \cite{amiriparian2024exhubert}. Such pre-trained models provide rich high-level speech representations. They capture prosodic and acoustic cues relevant to emotion and improve SER performance. Existing studies mainly use these models in two ways. The first is end-to-end fine-tuning for emotion classification. Wang et al.\  \cite{wang2021fine} report that fully fine-tuned wav2vec 2.0 and HuBERT reach about 73.01\% 
weighted accuracy on the popular IEMOCAP database in a speaker-independent setting, using 4 emotion classes (anger, happiness, sadness, and neutral). This shows the potential of end-to-end adaptation of pre-trained encoders to SER.

Another use of foundation models is to freeze the pre-trained encoder as a feature extractor and to train an independent downstream classifier only on its high-level speech representations. Pepino et al.\  \cite{pepino2021emotion} use multi-layer wav2vec 2.0 representations with a shallow network and surpass spectrogram-based and hand-crafted baselines on the IEMOCAP and RAVDESS databases. Furthermore, Emotion2Vec, with only a linear layer trained, outperforms mainstream general SSL models and emotion-specific models on IEMOCAP \cite{ma2023emotion2vec}. This indicates that emotion-oriented self-supervised representations are competitive in low-resource settings. In this work, we adopt the second approach.

\begin{figure*}[t]
  \centering
  \includegraphics[width=0.85\textwidth]{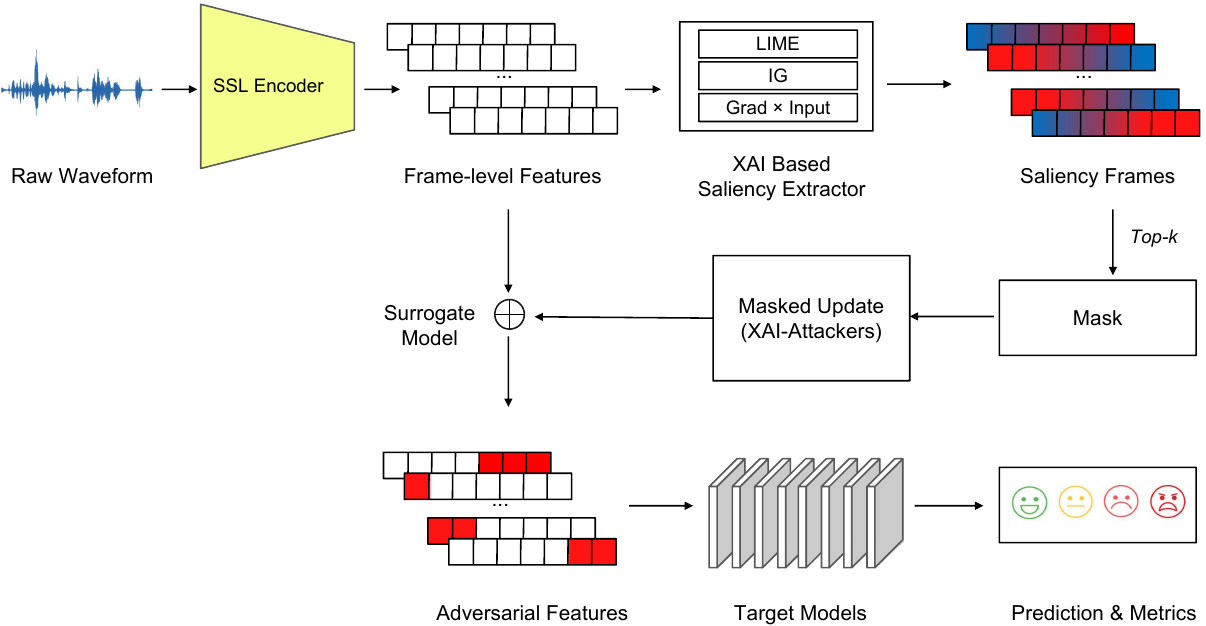}
  \caption{Workflow of SIGMA for SER. Raw waveforms are first encoded by a self-supervised speech encoder into frame-level features. An XAI saliency extractor on a surrogate classifier assigns per-frame importance. The top-$k$ salient coordinates define a binary mask under a fixed sparsity budget. Attacks then update only inside the mask under a norm budget for a small number of iterations, yielding adversarial features. The perturbation process is conducted in the feature space as a controlled analytical setting.}
  
  \vspace{-3mm}
  \label{fig:framework}
\end{figure*}

\subsection{Adversarial Attacks}
Adversarial attacks deceive deep neural networks (DNNs) via imperceptible perturbations, spanning white-box (e.g., FGSM~\cite{advexample2015}, PGD~\cite{madry2018towards}) and black-box settings~\cite{pmlr-v97-guo19a}. While traditional dense methods modify the entire input, sparse attacks (e.g., One-pixel attack~\cite{8601309}) selectively alter only a fraction of features. Crucially, sparse perturbations serve as diagnostic tools to pinpoint specific model vulnerabilities while remaining highly effective.

Adversarial attacks against audio systems can be broadly classified into two types: iterative gradient-based attacks and generator-based attacks. Iterative gradient-based techniques typically operate with a white-box setting, leveraging gradient information from the victim model to craft minimal perturbations that induce misclassification.

For instance, Neekhara et al.~\cite{NeekharaHPDMK19} discovered audio-agnostic universal perturbations by iteratively optimising the normalised Levenshtein distance to attack automatic speech recognition (ASR) systems. Similarly, Kim et al.~\cite{KIM2023109286} investigated the relationship between adversarial transferability and noise sensitivity, proposing a method that injects additive noise during gradient ascent to enhance transferability. In the domain of speaker verification, Zhang et al.~\cite{9413467} employed a Projected Gradient Descent (PGD) attack with momentum to generate effective, text-independent perturbations. However, a noticeable limitation of iterative methods is their computational expense, often requiring numerous iterations that can take hours to generate a single adversarial example. This high cost can hinder real-time application, and reducing iteration counts frequently compromises the attack success rate. Furthermore, adversarial examples crafted via these methods often exhibit poor transferability due to overfitting to the specific architecture of the target model.

To address these limitations, recent research has explored generator-based approaches, such as Generative Adversarial Networks (GANs)~\cite{goodfellow2020generative} and diffusion models~\cite{10889191}, which learn the underlying data distribution to produce more transferable perturbations in a single forward pass. Xie et al.~\cite{xie2021enabling} proposed a targeted attack framework that concatenates target class embeddings with intermediate generator features to attack various audio tasks. More recently, Chen et al.~\cite{10889191} leveraged a conditional diffusion model to reconstruct high-quality adversarial audio from an optimised Mel spectrogram, demonstrating superior attack success rate, transferability, and perceptual quality on speaker recognition systems. For black-box ASR scenarios, Yuan et al.~\cite{11023348} introduced a novel perturbation seed derived from a Formant Filter Bank, blending target command formants with original music features to create highly stealthy adversarial examples that are perceptually indistinguishable from benign audio.  However, these attacks lack the proper interpretation for the generated perturbations, leading to concerns especially for the reuse of the perturbations on privacy-sensitive data in the healthcare domain.

\subsection{Explainable Artificial Intelligence (XAI)}
XAI has advanced in deep learning. Common post hoc methods include Gradient×Input (GI), Integrated Gradients (IG), Layer-wise Relevance Propagation (LRP), Grad-CAM and its variants, and model agnostic methods such as LIME and SHAP \cite{sun2025explainable, shrikumar2017learning, li2025explainable}. GI multiplies the input by its gradient and serves as a first-order importance estimate \cite{shrikumar2017learning}. IG uses path integration and satisfies completeness and implementation invariance \cite{sundararajan2017axiomatic}. LRP provides a layer-wise relevance backpropagation framework \cite{montavon2019layer}. Grad-CAM localises evidence on feature maps \cite{selvaraju2017grad}. LIME and SHAP offer local surrogate and game theoretic views \cite{ribeiro2016should,lundberg2017unified}. Recent studies show that these XAI methods are effective on audio features \cite{akman2024audio}. Haunschmid et al.\  \cite{haunschmid2020audiolime} propose AudioLIME for music information retrieval. It first performs source separation to obtain interpretable components, then applies on and off perturbations to train a local linear surrogate and produces listenable local explanations. In SER, Nfissi et al.\  \cite{nfissi2024unveiling} use SHAP for feature importance and iterative feature enhancement, and report gains on the TESS and EMO-DB databases. This indicates that XAI can support feature selection and model analysis in SER.

There are two main approaches to combining XAI with adversarial methods. The first is to manipulate explanations: for example, the Focus-Shifting Attack \cite{huang2023focus} incorporates saliency maps into the loss, maintaining the original saliency distribution while shifting the explanation to evade detection. The goal is to mislead the explanation rather than align the original evidence. The second is to use explanations to guide or evaluate attacks. In audio tasks, recent research \cite{raina2024muting} on automatic speech recognition uses saliency analysis to characterise the impact of successful/failed samples on sensitive regions of the model, suggesting the feasibility of regionalised intervention based on attribution. There is also work proposing to use adversarial perturbations to evaluate attribution reliability and improve explanation evaluation protocols  \cite{nieradzik2025reliable}. However, systematic research on SER using ``explanation consistency'' as a generation constraint and parallel evaluation dimension remains scarce, which provides room for our approach.

\section{Method}

We propose a \textbf{S}al\textbf{I}ency-\textbf{G}uided sparse \textbf{M}ask \textbf{A}ttack (SIGMA) framework, which acts as a pluggable constraint mechanism. Its core idea is to first identify a small number of salient feature elements within the frame‑level representation most sensitive to the emotion‑recognition model using XAI techniques, and then inject small‑magnitude perturbations constrained by the $\ell_\infty$‑norm only within these selected elements. This approach improves explainability and energy concentration while maintaining competitive ASR under strict fairness, and enables efficient transfer by reusing masks computed on a surrogate.

\subsection{Problem Definition}

Given an original audio input $x$, it is encoded by a self-supervised speech representation model into a frame-level feature sequence of length $T$. Typically, each frame-level feature has $D$ dimensions:
\begin{equation}
F = \{ \mathbf{f}_t \in \mathbb{R}^{D} \mid t = 1, \ldots, T \}.
\end{equation}

The attacker targets a pre-trained SER model $f(\cdot)$.  
Under dual constraints of perturbation magnitude and sparsity, the goal is to construct an adversarial feature sequence:
\begin{equation}
\tilde{F} = F + \delta,
\end{equation}
such that:
\begin{equation}
f(\tilde{F}) \ne y,
\end{equation}
where $y$ is the ground-truth label of $x$.

\textit{Dual Constraints:}
\begin{itemize}
    \item Magnitude constraint: The perturbation magnitude on each feature dimension does not exceed a threshold $\varepsilon$:
    \begin{equation}
    \| \delta \|_\infty \le \varepsilon.
    \end{equation}


    \item Sparsity constraint: The perturbation \(\delta\) is only applied to a salient element set \(\Omega\) (determined by top-\(k\) saliency scores), where \(\Omega \subseteq \{1,\ldots,T\}\!\times\!\{1,\ldots,D\}\) and
\begin{equation}
\mathrm{supp}(\delta)\subseteq\Omega,\qquad |\Omega|=\lceil kTD\rceil,\quad k\in(0,1],
\end{equation}

    with $k$ being the predefined sparsity rate.
\end{itemize}

In summary, SIGMA follows the pipeline of ``saliency identification $\rightarrow$ sparse masking $\rightarrow$ attack optimisation”, and effectively disrupts SER model predictions while modifying only a tiny fraction of feature coefficients, revealing the model’s vulnerability to perturbations on semantically critical regions. We show the workflow in Fig.~\ref{fig:framework}.

\subsection{Feature Extraction}
The raw utterance \(x\) is first passed through a frozen self-supervised speech representation (SSL) encoder to obtain a frame-level representation $F$.

Adversarial perturbations are injected in this latent space in the later steps.  The choice of a frame‑level representation is motivated by three considerations.  First, modern SER pipelines typically adopt a two‑stage design: a frozen self‑supervised front‑end plus a lightweight classifier head. The encoder is often shared across downstream tasks; hence, the practitioner’s primary concern is how much the classifier relies on discriminative feature elements within each frame. Second, frame‑level features preserve temporal resolution, enabling attribution methods to identify a compact top subset of salient elements. Perturbing at this granularity confines noise to semantically critical regions, whereas crafting noise directly in the high‑dimensional waveform or spectrographic space would disperse over many samples and dilute the mask’s precision. This finer granularity concentrates perturbations on semantically critical regions and improves controllability. 
Third, compared to raw waveforms, \(F\) has far fewer degrees of freedom, which makes gradient estimation and \(\ell_\infty\)-norm updates practical; moreover, relative to Mel features, it is typically more semantically aligned with the task, which benefits saliency-guided sparsity.
\subsection{Saliency Identification}
To pinpoint the feature elements that most affect the model’s decision, we apply post‑hoc explainability methods to the feature sequence $F$.
Given the downstream SER classifier $f(\cdot)$, we derive a saliency map $S\!\in\!\mathbb{R}^{T\times D}$ whose entry $S_{t,d}$ quantifies the contribution of feature element $F_{t,d}$ to the logit of the ground-truth label $y$.

Given the downstream SER classifier $f(\cdot)$, we derive a saliency map $S\!\in\!\mathbb{R}^{T\times D}$ whose entry $S_{t,d}$ quantifies the contribution of element $F_{t,d}$ to the logit of the ground-truth label $y$. Specifically, we experiment with three representative methods: Gradient×Input (GI) \cite{shrikumar2017learning}, Integrated Gradients (IG) \cite{sundararajan2017axiomatic}, and LIME \cite{ribeiro2016should}.  
GI directly multiplies the input by the gradient of the target logit, while IG estimates attribution by accumulating gradients along a linear interpolation from a zero baseline to the input. LIME, in contrast, perturbs the input locally and fits a linear surrogate model to approximate frame‑level importance, after which the scalar score is broadcast to all $D$ dimensions of that frame.

We then use this saliency map $S$  to select the top‑$k$ elements as the masking target for perturbation.


\subsection{Sparse Masking}

Once the saliency map $S$ is obtained, SIGMA flattens $S$ and selects a salient element set $\Omega$ consisting of the indices of the top‑$k$ feature elements with the highest saliency scores. To enforce this sparsity constraint during perturbation, we define a binary mask matrix $M \in \{0,1\}^{T \times D}$ element-wise as
\begin{equation}
M_{t,d} = 
\begin{cases}
1 & \text{if } (t,d) \in \Omega,\\
0 & \text{otherwise.}
\end{cases}
\end{equation}

Here, $M_{t,d}=1$ explicitly denotes that the element $F_{t,d}$ will be perturbed, while $0$ means it remains unchanged.

This element‑level hard top‑$k$ mask serves two purposes. First, it enforces an explicit sparsity budget by ensuring that only a fixed proportion $k$ of the most salient coefficients is subject to perturbation regardless of the perturbation magnitude $\varepsilon$. Second, it preserves attribution faithfulness by confining modifications to those fine‑grained regions most critical to the classifier’s decision, thereby enhancing attack efficacy while reducing unintended distortion.

\begin{algorithm}[t]
\caption{SIGMA: Saliency-Guided Sparse Mask Attack (SIGMA-PGD$_{0}$ variant)}
\label{alg:sgsma}
\begin{algorithmic}[1]
\Require input $x$, SSL encoder $\mathcal{E}$, classifier $f$, ground-truth $y$, sparsity rate $k$, $\ell_\infty$ budget $\epsilon$, steps $N$, XAI method $\mathcal{A}$
\State $F \leftarrow \mathcal{E}(x)$ \Comment{frame-level features}
\State $S \leftarrow \mathcal{A}(f, F, y)$ \Comment{compute saliency map}
\State $\Omega \leftarrow \text{TopK\_indices}(S, k)$
\State $M \leftarrow \mathbb{I}[\Omega]$ \Comment{binary mask same shape as $F$}
\State $\delta \leftarrow \mathbf{0}_{T\times D}$
\For{$t = 1$ to $N$}
  \State $g \leftarrow \nabla_{F} L(f(F+\delta), y)$
  \State $\delta \leftarrow \text{clip}_{[-\epsilon,\epsilon]}\big( \delta + \alpha\cdot\text{sign}(g)\odot M \big)$ \Comment{$\alpha=\epsilon/N$}
\EndFor
\State \Return adversarial feature $F+\delta$
\end{algorithmic}
\end{algorithm}


\subsection{Masked Gradient Perturbation}
Having fixed the sparse binary mask $M$ that activates only the $K$ 
most salient feature elements, we optimise an adversarial perturbation within the masked feasible set using several first-order methods under the same sparsity and $\ell_\infty$ constraints.

\textbf{SIGMA}-PGD$_{0}$. Starting from $\delta^{(0)}=\mathbf{0}$, we perform masked signed-gradient updates with box projection:
\begin{equation}
\delta^{(t+1)}
=\operatorname{clip}\!\Bigl(\,\delta^{(t)}
+\alpha\,\operatorname{sgn}\!\bigl(\nabla_{F}\mathcal{L}(f(F+\delta^{(t)}),y)\bigr)\odot M\,\Bigr),
\end{equation}
where $\alpha=\varepsilon/N$ is the step size for $N$ iterations, and clips each entry to $[-\varepsilon,\varepsilon]$; because the update is masked by $M$, off-mask entries remain zero at all steps.
The final adversarial feature is $\tilde{F}=F+\delta^{(N)}$. Because $M$ is data-driven and input-specific, the same mask-guided scheme applies to transfer: $M$ is computed once on a surrogate via XAI and then reused on target models. Algorithm \ref{alg:sgsma} shows the workflow of SIGMA-PGD$_{0}$.

\textbf{SIGMA}-FW-$\ell_{1}$. We use the same mask $M$, per-coordinate cap $\varepsilon$, and aligned $\ell_{1}$ budget $\tau_{\ell_1}=\varepsilon K$ as in SIGMA-PGD$_{0}$. Let $F$ be the clean feature and $F^{(t)}$ the iterate. Define
\begin{equation}
m^{(t)} \;=\; \nabla_{F}\!\big(-\mathcal{L}(f(F^{(t)}),y)\big).
\end{equation}
Over the masked feasible set $\mathcal{S}_M$, the linear oracle constructs a sparse $\delta^{\star}$ inside $M$ and sets $v^{(t)}=F-\delta^{\star}$. Sort indices in $\Omega$ by decreasing $|m^{(t)}_i|$, let
\begin{equation}
K_b \;=\; \min\!\Big(|\Omega|,\ \big\lceil\tau_{\ell_1}/\varepsilon\big\rceil\Big), 
\qquad
r \;=\; \tau_{\ell_1}-\varepsilon\,(K_b-1)\in[0,\varepsilon],
\end{equation}
and define
\begin{equation}
\delta^{\star}_i \;=\;
\begin{cases}
\varepsilon\,\mathrm{sgn}\!\big(m^{(t)}_i\big), & \text{for the first } K_b-1 \text{ indices},\\[2pt]
r\,\mathrm{sgn}\!\big(m^{(t)}_i\big),          & \text{for the }K_b\text{-th index},\\[2pt]
0,                                              & \text{otherwise,}
\end{cases}
\quad (i\in\Omega).
\end{equation}
The update is
\begin{equation}
F^{(t+1)} \;=\; (1-\gamma_t)\,F^{(t)} \;+\; \gamma_t\,v^{(t)}, 
\qquad \gamma_t\in(0,1].
\end{equation}
This matches the implementation: selection and capping occur only on $\Omega$; at most $K_b=\lceil\tau_{\ell_1}/\varepsilon\rceil$ coordinates are non-zero (the last takes the remainder); the iterate remains within the masked $\ell_1\cap\ell_\infty$ budget. A momentum buffer $m^{(t)}\!\leftarrow\!\beta m^{(t-1)}+(1-\beta)\nabla_{F}(-\mathcal{L})$ may replace $m^{(t)}$ in the sort.

\textbf{SIGMA}-Sparsefool. We keep the same mask $M$, the same $K$, and the same budgets as above. At iterate $\delta^{(t)}$ and for each $k\neq y$, define
\begin{equation}
w_k \;=\; \nabla_{F} f_k\!\big(F+\delta^{(t)}\big)\;-\;\nabla_{F} f_y\!\big(F+\delta^{(t)}\big),
\end{equation}
\begin{equation}
\rho_k \;=\; f_y\!\big(F+\delta^{(t)}\big)\;-\;f_k\!\big(F+\delta^{(t)}\big).
\end{equation}
On the mask support, allocate an $\ell_1$ step with per-coordinate caps by solving
\begin{equation}
\begin{aligned}
a_k^\star \;=\; \arg\min_{a\ge 0}\ \|a\|_{1}
\quad\text{s.t.}\quad \langle |w_k|, a\rangle \ge \rho_k,\\
0\le a \le \varepsilon M,\quad \|a\|_{1}\le \tau_{\text{rem}}.
\end{aligned}
\end{equation}
Pick $k^\star=\arg\min_k \|a_k^\star\|_{1}$ among feasible classes (otherwise choose the one with the largest budget-limited progress), and update with masking and box projection:
\begin{equation}
\delta^{(t+1)} \;=\; 
\operatorname{clip}_{[-\varepsilon,\varepsilon]}\!\Big(
\delta^{(t)} \;+\; \operatorname{sgn}(w_{k^\star}) \odot a_{k^\star}^\star
\Big).
\end{equation}
Stop when misclassified or when no progress is made. The fixed, input-specific mask enforces sparsity and supports transfer by reusing it on target models.

\begin{table}[h]
\centering
\caption{Data distribution of the IEMOCAP and TESS datasets.}
\begin{tabular}{lcccc}
\toprule
\textbf{Dataset} & \textbf{Train} & \textbf{Valid} & \textbf{Test} & \textbf{Total} \\
\midrule
IEMOCAP (4-class) & 3,556 & 890 & 1,085 & 5,531 \\
TESS (7-class)    & 1,260 & 140 & 1,400  & 2,800 \\
\bottomrule
\end{tabular}
\label{tab:dataset-summary}
\end{table}

\subsection{Evaluation Metrics}

To assess each adversarial strategy under a given perturbation budget~$\varepsilon$, we report metrics spanning \emph{attack effectiveness}, \emph{explanation consistency}, and \emph{computational cost}. Let $\mathbf{s}^{c},\mathbf{s}^{a}\in\mathbb{R}^{T}$ denote element-level saliency scores for the clean and adversarial inputs, both normalised to $[0,1]$.

\paragraph{Attack Effectiveness}
\begin{itemize}
    \item \textbf{Clean Accuracy.} The classification unweighted accuracy (UA) of the model on unperturbed inputs, used as the reference baseline.

    \item \textbf{Attack Success Rate (ASR).}
          The proportion of originally correctly classified inputs that are misclassified after the perturbation is applied.

    \item \textbf{Sparsity.}
    The proportion of perturbed entries over the frame domain grid. 
    \textbf{\item \textbf{$L_2$ Avg.} }
    For each sample $i$, we compute the RMS $\ell_2$ norm of $\delta_i$ over its valid (non-padded) entries and then average over the set.

\end{itemize}

\paragraph{Explanation Consistency}
\begin{itemize}
           \item \textbf{Top‑$k$ Intersection ($\mathrm{Top\!-\!k}\cap$).}  
          Let $S^{c}$ and $S^{a}$ be the index sets of the $k$ most salient elements from $\mathbf{s}^{c}$ and $\mathbf{s}^{a}$.  We compute
          \[
              \frac{|S^{c}\cap S^{a}|}{k}\in[0,1].
          \]
          
          A higher value means the same elements remain salient after the attack.

    \item \textbf{Kendall’s $\boldsymbol{\tau}$.}
    The rank correlation between $\mathbf{s}^{c}$ and $\mathbf{s}^{a}$, measured using Kendall’s $\tau$ coefficient. A value close to 1 indicates strong agreement in the relative ordering of element‑level saliency.

    \item \textbf{$\boldsymbol{\Delta \text{Sal}}$ (Total Variation Distance).}
    After normalising saliency vectors $\mathbf{s}^{c}$ and $\mathbf{s}^{a}$ into probability distributions $\mathbf{p}$ and $\mathbf{q}$ (i.e., $\mathbf{p} = \mathbf{s}^{c} / \sum_t s_t^c$), compute:
    \[
        \Delta\text{Sal} = \|\mathbf{p} - \mathbf{q}\|_{1} \in [0,2].
    \]

    This metric reflects the total variation distance between clean and adversarial saliency distributions. Smaller values indicate better preservation of global attribution patterns.


\end{itemize}

\paragraph{Computational Cost}
\begin{itemize}
    \item \textbf{Time.} Average per-example attack time over successful cases (excludes XAI precomputation).
    \item \textbf{Saliency Time.} One-off XAI mask precomputation time (reported as total in all samples), which is reused across targets and not counted in attack rows.
\end{itemize}

\section{Experiments}
\subsection{Datasets}
To evaluate the generalisability and robustness of our adversarial attack strategy, we conduct experiments on two widely used SER datasets: IEMOCAP and TESS. Table \ref{tab:dataset-summary} reports the number of samples in each split.

\textbf{IEMOCAP.}
The Interactive Emotional Dyadic Motion Capture (IEMOCAP) corpus \cite{busso2008iemocap} contains 5531 audio utterances from 10 professional actors (5 male, 5 female) recorded over five dyadic sessions. Each utterance is labelled with categorical and dimensional emotion annotations. We adopt a 4-class setting commonly used in SER studies, where happy and excited are merged, alongside neutral, angry, and sad. Following the dataset split strategy of other SER researches \cite{ma2023emotion2vec, zhao2021self}, we divide the data into five folds of approximately equal size and report results on one held-out fold (Fold 1 in our experiments), with the remaining samples split into training (80\%) and validation (20\%) sets.

\textbf{TESS.} The Toronto Emotional Speech Set (TESS) \cite{pichora2020toronto} comprises 2800 utterances recorded by two female speakers reading 200 target words under seven emotional conditions. The standard 7-class scheme is used: anger, disgust, fear, happiness, neutral, pleasant surprise, and sadness. To avoid the impact of speaker characteristics, we use a leave-one-speaker-out (LOSO) protocol: in each fold, one speaker is held out for testing (OAF or YAF). The remaining speaker forms the training pool, from which we draw a stratified 10\% split for validation, with the rest used for training. The random seed is used for reproducibility. We then report test results for the held-out speaker. Final performance is the average over both folds. 

\begin{table}[htbp]
  \caption{Classification baseline across the clean SSL features and datasets.}
  \label{tab:clean_baseline}
  \centering
  \begin{tabular}{l l c c c}
    \toprule
    \multirow{2}{*}{SSL} & \multirow{2}{*}{Dataset} & \multicolumn{3}{c}{Unweighted Accuracy (UA, \%)} \\
    \cmidrule(lr){3-5}
     &  & Zhao19 & BaseModel  & Emo18 \\
    \midrule
    \multirow{2}{*}{Emotion2Vec} & IEMOCAP & 71.56& 75.58  & 70.93 \\
                                 & TESS    & 77.64 & 69.86 &63.79  \\
    \midrule
    \multirow{2}{*}{WavLM}       & IEMOCAP & 69.06 &68.39 & 68.19 \\
                                 & TESS    & 89.36 & 73.64 & 70.00  \\
    \midrule
    \multirow{2}{*}{HuBERT}      & IEMOCAP &  66.19&  67.44& 66.20 \\
                                 & TESS    & 64.36 & 57.79 & 67.14 \\
    \bottomrule
  \end{tabular}
\end{table}

\subsection{Speech Representations}
We evaluate three widely used self-supervised speech encoders and use their released checkpoints as frozen feature extractors. We only consume their frame-level features without any further fine-tuning.

\textbf{Emotion2vec}
\cite{ma2023emotion2vec} is a Transformer-based SSL encoder that is first self-supervised pre-trained and then emotion-oriented fine-tuned on large mixed corpora; we use the public \textit{emotion2vec + base} checkpoint and  adopt it as a frozen feature extractor with per-frame dimensionality $D{=}768$.

\textbf{WavLM} \cite{wang2023speech} is a wav2vec 2.0 style masked-prediction Transformer for speech. We adopt the checkpoint \textit{speechbrain/emotion-diarization-wavlm-large}, which is fine-tuned for emotion diarisation
across multiple datasets. The length of extracted  frame-level features  is  $D{=}1024$.

\textbf{HuBERT}
\cite{yang2021superb} is a BERT-style masked-prediction speech encoder that predicts hidden units obtained via offline clustering. We adopt the checkpoint \textit{superb/hubert-base-superb-er}, which is fine-tuned for the SUPERB Emotion Recognition task, and use its frame-level features with dimensionality $D{=}768$.


\subsection{Downstream Classifiers}
We implement three classifier architectures with distinct inductive biases: a two-layer MLP (\textbf{BaseModel}), a 1-D CNN (\textbf{Zhao19}), and a shallow CNN (\textbf{Emo18}). Each consumes the frame-level SSL representations introduced in the previous section. We train and evaluate all $3\times 3$ encoder-classifier combinations under identical optimisation settings and data splits for fair comparison.

\begin{itemize}
    \item \textbf{BaseModel} \cite{ma2023emotion2vec}: A simple feedforward neural network with two fully connected layers and a ReLU activation in between. Each frame is first projected to 256 dimensions. Mask-aware mean pooling is applied across time, followed by a linear output layer.
    \item \textbf{Zhao19} \cite{zhao2019speech}: Consists of four 1D convolutional layers with increasing channel sizes. Each layer is followed by max pooling and an ELU activation. The output is aligned to the input mask using nearest-neighbour interpolation and aggregated by mask-aware pooling before classification.
    \item \textbf{Emo18} \cite{tzirakis2018end}: Uses larger convolutional kernels (8/6/6) and stronger downsampling (stride 10/8/8) to extract more abstract but temporally coarse emotional features. Each layer includes BatchNorm and ReLU activations.
\end{itemize}

\begin{table*}[htbp]
  \centering
  \small
  \renewcommand{\arraystretch}{0.9}
  \caption{Sparse white-box comparison under the same budget on IEMOCAP. ``Avg time'' is computed over successful samples only.}

  \label{tab:m1-iemocap}
  \resizebox{0.7\linewidth}{!}{
  \begin{tabular}{lllccc}
    \toprule
    \textbf{SSL} & \textbf{Model} & \textbf{Attack} & \textbf{ASR (\%)} $\uparrow$ & \textbf{Avg time (s)} $\downarrow$ & \textbf{Sparsity} $\downarrow$ \\
    \midrule
    \multirow{6}{*}{\makecell[c]{Emotion2Vec}} 
      & \multirow{6}{*}{basemodel} 
        & PGD$_{0}$            &\textbf{67.13} & .0069 & .1563 \\
      &                  & \textbf{SIGMA-PGD$_{0}$}      & 64.87 & \textbf{.0060} & \textbf{.1562 }\\
      \cmidrule(lr){3-6}
      &                  & FW-$\ell_{1}$           & \textbf{69.25} & .0472 & .1664 \\
      &                  & \textbf{SIGMA-FW-$\ell_{1}$}     & 66.25 & \textbf{.0467 }& \textbf{.1572 }\\
      \cmidrule(lr){3-6}
      &                  & Sparsefool      & \textbf{61.48} & .0315 & \textbf{.0188} \\
      &                  & \textbf{SIGMA-Sparsefool} & 59.47& \textbf{.0314} & .0200\\
    \midrule
    \multirow{6}{*}{\makecell[c]{WavLM}} 
      & \multirow{6}{*}{zhao19} 
        & PGD$_{0}$            & \textbf{75.66} & .0110 & .1964 \\
      &                  & \textbf{SIGMA-PGD$_{0}$}      & 69.48 & \textbf{.0095 }& \textbf{.1963 }\\
      \cmidrule(lr){3-6}
      &                  & FW-$\ell_{1}$           & \textbf{63.43} & .0554 & .1744 \\
      &                  & \textbf{SIGMA-FW-$\ell_{1}$}     & 60.62 & \textbf{.0515} & \textbf{.1572 }\\
      \cmidrule(lr){3-6}
      &                  & Sparsefool      & \textbf{58.65} & .0445& \textbf{.0586} \\
      &                  & \textbf{SIGMA-Sparsefool} & 56.26 & \textbf{.0423 }& .0597 \\
    \midrule
    \multirow{6}{*}{\makecell[c]{HuBERT}} 
      & \multirow{6}{*}{basemodel} 
        & PGD$_{0}$            &\textbf{95.22 }& .0074 & .1564 \\
      &                  & \textbf{SIGMA-PGD$_{0}$}      & 93.82 & \textbf{.0065 }& \textbf{.1563} \\
      \cmidrule(lr){3-6}
      &                  & FW-$\ell_{1}$           & \textbf{97.17} & .0344 & .1642 \\
      &                  & \textbf{SIGMA-FW-$\ell_{1}$}     & 95.30 & \textbf{.0334} & \textbf{.1557} \\
      \cmidrule(lr){3-6}
      &                  & Sparsefool      & \textbf{52.67} & .0325 & \textbf{.0216} \\
      &                  & \textbf{SIGMA-Sparsefool} & 50.28 &\textbf{.0322}& .0233 \\
    \bottomrule
  \end{tabular}
  }
\end{table*}

\begin{table*}[htbp]
  \centering
  \small
  \renewcommand{\arraystretch}{0.9}
  \caption{Sparse white-box comparison under the same budget on TESS. ``Avg time'' is computed over successful samples only; Saliency precompute time for Ours is reported elsewhere.}
  \label{tab:m1-tess}
  \resizebox{0.7\linewidth}{!}{
  \begin{tabular}{lllccc}
    \toprule
    \textbf{SSL} & \textbf{Model} & \textbf{Attack} & \textbf{ASR (\%)} $\uparrow$ & \textbf{Avg time (s)} $\downarrow$ & \textbf{Sparsity $\downarrow$ } \\
    \midrule
    \multirow{6}{*}{\makecell[c]{Emotion2Vec}} 
      & \multirow{6}{*}{zhao19} 
        & PGD$_{0}$            &\textbf{99.17} & .0015 &.2571 \\
      &                  & \textbf{SIGMA}-PGD$_{0}$      & 95.86  & \textbf{.0011} & \textbf{.2459 }\\
      \cmidrule(lr){3-6}
      &                  & FW-$\ell_{1}$           & \textbf{89.77} &.0078&.2416  \\
      &                  & \textbf{SIGMA-FW-$\ell_{1}$}     &71.09  & \textbf{.0076 }&\textbf{.1999} \\
      \cmidrule(lr){3-6}
      &                  & Sparsefool      & 79.74 & .0136 & \textbf{.0214} \\
      &                  & \textbf{SIGMA-Sparsefool} &\textbf{81.68} & \textbf{.0116} &.0237 \\
    \midrule
    \multirow{6}{*}{\makecell[c]{WavLM}} 
      & \multirow{6}{*}{zhao19} 
        & PGD$_{0}$            & \textbf{66.10} & .0017 & .2224 \\
      &                  & \textbf{SIGMA-PGD$_{0}$}      & 60.87 & \textbf{.0013 }& \textbf{.2219}\\
      \cmidrule(lr){3-6}
      &                  & FW-$\ell_{1}$           & \textbf{58.77} & .0101 &.2752 \\
      &                  & \textbf{SIGMA-FW-$\ell_{1}$}     & 49.52 & \textbf{.0100} & \textbf{.1999}\\
      \cmidrule(lr){3-6}
      &                  & Sparsefool      & \textbf{55.15} &.0433 & \textbf{.0820} \\
      &                  & \textbf{SIGMA-Sparsefool} & 50.16 & \textbf{.0408}&.0883 \\
    \midrule
    \multirow{6}{*}{\makecell[c]{HuBERT}} 
      & \multirow{6}{*}{emo18} 
        & PGD$_{0}$            &\textbf{95.10}&.0015  &.2492\\
      &                  & \textbf{SIGMA-PGD$_{0}$}      & 91.18 &\textbf{.0012} & \textbf{.2410} \\
      \cmidrule(lr){3-6}
      &                  & FW-$\ell_{1}$           & \textbf{94.34} & .0085 & .2933 \\
      &                  & \textbf{SIGMA-FW-$\ell_{1}$}     & 83.68 & \textbf{.0084} & \textbf{.1999} \\
      \cmidrule(lr){3-6}
      &                  & Sparsefool      & \textbf{81.72} &.0156  & \textbf{.0095} \\
      &                  & \textbf{SIGMA-Sparsefool} &80.52 &\textbf{.0132}&.0102 \\
    \bottomrule
  \end{tabular}
  }
\end{table*}

During the training stage, the loss is cross-entropy. The optimiser is RMSprop with learning rate \(5\times 10^{-4}\) and momentum \(0.9\). Weight decay is \(1\times 10^{-5}\). The scheduler is CyclicLR with base learning rate \(5\times 10^{-4}\), maximum learning rate \(10^{-3}\), and step\_size\_up = 10. Training runs for 100 epochs. For attribution, GI computes the absolute gradient times input \( \lvert \nabla_{\mathbf{x}}\mathcal{L}\odot\mathbf{x} \rvert \);
IG uses a zero baseline with \(S_{\mathrm{IG}}=50\) steps;
LIME adopts frame-wise grouping with \(N_{\mathrm{LIME}}=20\) perturbation samples.

Experiments are performed on a single NVIDIA A30 GPU. 
All models achieve comparable performance on clean (non-adversarial) data, establishing a fair baseline for subsequent adversarial perturbation experiments. Table~\ref{tab:clean_baseline} reports the clean unweighted accuracy (UA) for all encoder–classifier combinations considered.  In the experimental stage, we select the best performing SSL + classifier combination to present.

\subsection{White- and Black-box Experiments }

\begin{table*}[t]
\centering
\small
\caption{\textbf{ White-box sparse PGD$_{0}$ shift on Emotion2Vec on IEMOCAP}
(baseline = PGD$_{0}$, $\varepsilon=0.02$, $K=20\%$, steps$=10$).
Metrics (mean$\pm$std) on target clean-correct samples with target-side saliency. $^{*}$ denote significant differences vs baseline by paired test ($p{<}0.001$)  } 

\resizebox{0.7\linewidth}{!}{
\begin{tabular}{l l l c c}
\toprule
\textbf{Target} & \textbf{Attack} &\textbf{ TopK$\cap$ $\uparrow$} &\textbf{ $\tau_b$ $\uparrow$} &\textbf{ $\Delta$Sal $\downarrow$} \\
\midrule
\multirow{4}{*}{BaseModel}
 & PGD$_{0}$                 & 0.8575$\pm$0.0369 & 0.7483$\pm$0.0485 & 0.2962$\pm$0.1308 \\
 & GI-PGD$_{0}$                  & \textbf{0.8606}$\pm$0.0382$^{*}$ & \textbf{0.7550}$\pm$0.0523$^{*}$ & \textbf{0.2921}$\pm$0.1307$^{*}$ \\
 &    IG-PGD$_{0}$               & \textbf{0.8627}$\pm$0.0396$^{*}$ & \textbf{0.7584}$\pm$0.0550$^{*}$ & \textbf{0.2901}$\pm$0.1306$^{*}$ \\
 & LIME-PGD$_{0}$     & \textbf{0.8726}$\pm$0.0461$^{*}$ & \textbf{0.7714}$\pm$0.0653$^{*}$ & \textbf{0.2792}$\pm$0.1299$^{*}$ \\
\midrule
\multirow{4}{*}{Zhao19}
 & PGD$_{0}$      & 0.7372$\pm$0.0622 & 0.6098$\pm$0.0618 & 0.3767$\pm$0.1511 \\
 & GI-PGD$_{0}$                  & \textbf{0.7437}$\pm$0.0625$^{*}$ & \textbf{0.6180}$\pm$0.0632$^{*}$ & \textbf{0.3760}$\pm$0.1522$^{*}$ \\
 & IG-PGD$_{0}$                  & \textbf{0.7496}$\pm$0.0657$^{*}$ & \textbf{0.6250}$\pm$0.0676$^{*}$ & \textbf{0.3755}$\pm$0.1532$^{*}$ \\
 & LIME-PGD$_{0}$               & \textbf{0.7668}$\pm$0.0746$^{*}$ & \textbf{0.6458}$\pm$0.0799$^{*}$ & \textbf{0.3721}$\pm$0.1590$^{*}$ \\
\midrule
\multirow{4}{*}{Emo18}
 & PGD$_{0}$                 & 0.7285$\pm$0.0706 & 0.6016$\pm$0.0696 & 0.3739$\pm$0.1550 \\
 & GI-PGD$_{0}$                  & \textbf{0.7375}$\pm$0.0716$^{*}$ & \textbf{0.6125}$\pm$0.0718$^{*}$ & \textbf{0.3724}$\pm$0.1569$^{*}$ \\
 & IG-PGD$_{0}$                  & \textbf{0.7416}$\pm$0.0728$^{*}$ & \textbf{0.6173}$\pm$0.0738$^{*}$ & \textbf{0.3715}$\pm$0.1579$^{*}$ \\
 & LIME-PGD$_{0}$               & \textbf{0.7636}$\pm$0.0835$^{*}$ & \textbf{0.6442}$\pm$0.0887$^{*}$ & \textbf{0.3656}$\pm$0.1655$^{*}$ \\
\bottomrule
\end{tabular}}
\label{tab:whitebox_drift_emotion2vec_noaopc}
\end{table*}

\begin{table*}[t]
\centering
\small
\caption{\textbf{ White-box sparse PGD$_{0}$ shift on Emotion2Vec on TESS. (baseline = PGD$_{0}$, $\varepsilon=0.02$, $K=20\%$, steps$=10$).
Metrics (mean$\pm$std) on target clean-correct samples with target-side saliency. $^{*}$ denote significant differences vs baseline by paired test ($p{<}0.001$)}}
\resizebox{0.7\linewidth}{!}{
\begin{tabular}{l l l c c}
\toprule
\textbf{Target} & \textbf{Attack} & \textbf{TopK$\cap$ $\uparrow$} & \textbf{$\tau_b$ $\uparrow$} & \textbf{$\Delta$Sal $\downarrow$} \\
\midrule
\multirow{4}{*}{BaseModel}
 & PGD$_{0}$                 & 0.9342$\pm$0.0255 & 0.8719$\pm$0.0433 &  0.2130$\pm$0.1068 \\
 & GI-PGD$_{0}$                  & 
\textbf{0.9414}$\pm$0.0212$^{*}$ & \textbf{0.8930}$\pm$0.0392$^{*}$ & \textbf{0.1844}$\pm$0.0939$^{*}$ \\
 &    IG-PGD$_{0}$               & \textbf{0.9524}$\pm$0.0189$^{*}$ & \textbf{0.9094}$\pm$0.0344$^{*}$ & \textbf{0.1592}$\pm$0.0819$^{*}$ \\
 & LIME-PGD$_{0}$     & \textbf{0.9637}$\pm$0.0187$^{*}$ & \textbf{0.9240}$\pm$0.0346$^{*}$ & \textbf{0.1240}$\pm$0.0620$^{*}$ \\
\midrule
\multirow{4}{*}{Zhao19}
 & PGD$_{0}$      & 0.8632$\pm$0.0242 &  0.7666$\pm$0.0339 &  0.2559$\pm$0.0961\\
 & GI-PGD$_{0}$                  & \textbf{0.8778}$\pm$0.0213$^{*}$ & \textbf{0.7909}$\pm$0.0316$^{*}$ & \textbf{0.2342}$\pm$0.0878$^{*}$ \\
 & IG-PGD$_{0}$                  & \textbf{0.8890}$\pm$0.0274$^{*}$ & \textbf{0.8063}$\pm$0.0421$^{*}$ & \textbf{0.2163}$\pm$0.0801$^{*}$ \\
 & LIME-PGD$_{0}$               & \textbf{0.9081}$\pm$0.0179$^{*}$ & \textbf{0.8298}$\pm$0.0280$^{*}$ & \textbf{0.1715}$\pm$0.0590$^{*}$ \\
\midrule
\multirow{4}{*}{Emo18}
 & PGD$_{0}$                 & 0.8951$\pm$0.0353 &  0.8065$\pm$0.0496 & 0.2553$\pm$0.1368 \\
 & GI-PGD$_{0}$                  & \textbf{0.9055}$\pm$0.0342$^{*}$ & \textbf{0.8273}$\pm$0.0504$^{*}$ & \textbf{0.2332}$\pm$0.1274$^{*}$ \\
 & IG-PGD$_{0}$                  & \textbf{0.9140}$\pm$0.0362$^{*}$ & \textbf{0.8395}$\pm$0.0531$^{*}$ & \textbf{0.2179}$\pm$0.1209$^{*}$ \\
 & LIME-PGD$_{0}$               & \textbf{0.9278}$\pm$0.0338$^{*}$ & \textbf{0.8588}$\pm$0.0517$^{*}$ & \textbf{0.1792}$\pm$0.0984$^{*}$ \\
\bottomrule
\end{tabular}}
\label{tab:whitebox_drift_emotion2vec_noaopc_tess}
\end{table*}

We compare three SSL representations and three sparse attacks (PGD$_{0}$, Frank–Wolfe, Sparsefool) in a white-box setting SIGMA on two datasets. To ensure strict fairness, all methods use the same epsilon, the same top-$k$ sparsity budget, and the same number of iterations. The SIGMA variants update only within a fixed mask $M$. All other hyper-parameters remain unchanged. Results on the IEMOCAP and TESS databases are reported in Tables \ref{tab:m1-iemocap} and \ref{tab:m1-tess}.

On IEMOCAP, the trends are consistent across methods under the same $epsilon$, top-$k$ budget, and iteration count. For PGD$_{0}$, SIGMA-PGD$_{0}$ shows equal or lower average generation time in all three settings (Emotion2Vec: .0060 vs .0069; WavLM: .0095 vs .0110; HuBERT: .0065 vs .0074). ASR is typically lower than the baseline by about 1–6 percentage points. For FW-$\ell_{1}$, SIGMA-FW-$\ell_{1}$ also tends to reduce time and often yields lower sparsity. This suggests a more concentrated use of the same $\ell\_1$ budget on salient support, with a small drop in ASR. For Sparsefool, time is similar or slightly better and ASR is slightly lower. The sparsity is marginally higher in all three settings (+0.0011 to +0.0017). This likely reflects more coordinates touching very small per-coordinate caps under the $\ell_\infty$ constraint. The values remain comparable within the unified budget.  

Under matched budgets, we observe similar trends on TESS. For PGD$_{0}$, SIGMA-PGD$_{0}$ reduces average generation time in all three settings (Emotion2Vec: $0.0011$ vs $0.0015$; WavLM: $0.0013$ vs $0.0017$; HuBERT: $0.0012$ vs $0.0015$). ASR drops modestly ($\approx 3$--$5$ pp), while sparsity is equal or lower ($-0.0112/-0.0005/-0.0082$). For FW-$\ell_{1}$, SIGMA-FW-$\ell_{1}$ keeps time essentially unchanged and consistently lowers sparsity to $\approx 0.20$, with a larger ASR reduction ($\approx 9$--$19$ pp). For Sparsefool, time is similar or better, and ASR is slightly lower in two settings and higher in one (Emotion2Vec: $+1.94$ pp). Sparsity is marginally higher across all three ($+0.0023$, $+0.0063$, $+0.0007$), indicating more coordinates touching very small per-coordinate caps under the $\ell_\infty$ constraint. Overall, the TESS results echo IEMOCAP: SIGMA variants tend to reduce time and concentrate perturbations, reflecting a clear trade-off where a portion of ASR (e.g., for $FW-l_{1}$) is willingly sacrificed for improved sparsity control and explanation consistency.

It is worth noting that "Avg time" purely reflects the online attack generation and does not include the one-off SIGMA mask pre-computation. To avoid overstating efficiency, we must explicitly discuss the total latency" (pre-computation + attack speed) for a single-sample attack scenario. For a completely new input, the initial latency required to compute the XAI mask is prohibitive for real-time applications. Consequently, the Total Latency of SIGMA in a single-target, single-sample white-box setting is higher than that of standard iterative baselines.

However, the efficiency of SIGMA manifests in its amortised cost. The data-driven mask is reusable across different attack methods and target models sharing the same SSL front-end. Therefore, in cross-model transfer scenarios or multi-target evaluations, the high initial Total Latency is heavily amortised. Furthermore, masks can be computed offline and enable rapid online adversarial generation for high-throughput use.

\begin{table}[ht]
\centering
\caption{WHITE-BOX TRANSFER RESULTS ON IEMOCAP. BASELINE RESULTS REPRESENT DIRECT TARGET-SPECIFIC WHITE-BOX ATTACKS (SERVING AS UPPER BOUNDS), AS THEIR SPARSE SELECTION INDICES ARE HIGHLY TARGET-DEPENDENT.}
\label{tab:transfer}

\resizebox{\linewidth}{!}{
\begin{tabular}{l lcccc}
\toprule
\multirow{3}{*}{\textbf{SSL \& Surrogate}} & \multirow{3}{*}{\textbf{Attacker}} & \multicolumn{4}{c}{\textbf{Target model}} \\
\cmidrule(lr){3-6}
& & \multicolumn{2}{c}{Zhao19} & \multicolumn{2}{c}{Emo18} \\
\cmidrule(lr){3-4}\cmidrule(lr){5-6}
& & ASR (\%) $\uparrow$ & Speed (s) $\downarrow$ & ASR (\%) $\uparrow$ & Speed (s) $\downarrow$ \\
\midrule
\multirow{6}{*}{\makecell[c]{Emotion2Vec\\ +\\ basemodel}}
  & PGD$_{0}$               & \textbf{98.79} & .0059 & \textbf{94.28} & .0065 \\
  & \textbf{SIGMA-PGD$_{0}$} & 94.63 & \textbf{.0049} & 88.42 & \textbf{.0055} \\
  \cmidrule(lr){2-6}
  & FW-$\ell_{1}$            & \textbf{99.46} & .4916 & \textbf{95.23} & .4918 \\
  & \textbf{SIGMA-FW-$\ell_{1}$} & 96.24 & \textbf{.4910} & 89.78 & \textbf{.4914} \\
  \cmidrule(lr){2-6}
  & Sparsefool        & \textbf{73.46} & .0462 & \textbf{99.45} & .0510 \\
  & \textbf{SIGMA-Sparsefool} & 72.25 & \textbf{.0417} & 98.50 & .0512 \\
\bottomrule
\end{tabular}
}

\vspace{1.2ex}

\resizebox{\linewidth}{!}{
\begin{tabular}{l lcccc}
\toprule
\multirow{3}{*}{\textbf{SSL \& Surrogate}} & \multirow{3}{*}{\textbf{Attacker}} & \multicolumn{4}{c}{\textbf{Target model}} \\
\cmidrule(lr){3-6}
& & \multicolumn{2}{c}{BaseModel} & \multicolumn{2}{c}{Emo18} \\
\cmidrule(lr){3-4}\cmidrule(lr){5-6}
& & ASR (\%) $\uparrow$ & Speed (s) $\downarrow$ & ASR (\%) $\uparrow$ & Speed (s) $\downarrow$ \\
\midrule
\multirow{6}{*}{\makecell[c]{WavLM\\ +\\ zhao19}}
  & PGD$_{0}$               & 51.39 & .0058 & \textbf{78.30} & .0085 \\
  & \textbf{SIGMA-PGD$_{0}$} & 45.00 & \textbf{.0049} & 68.29 & \textbf{.0073} \\
  \cmidrule(lr){2-6}
  & FW-$\ell_{1}$            & \textbf{52.78} & .0343 & \textbf{73.99} & .0425 \\
  & \textbf{SIGMA-FW-$\ell_{1}$} & 45.69 & \textbf{.0272} & 64.26 & \textbf{.0359} \\
  \cmidrule(lr){2-6}
  & Sparsefool        & \textbf{17.08} & .1646 & \textbf{68.70} & .1419 \\
  & \textbf{SIGMA-Sparsefool} & 14.31 & .1743 & 60.65 & .1578 \\
\bottomrule
\end{tabular}
}

\vspace{1.2ex}

\resizebox{\linewidth}{!}{
\begin{tabular}{l lcccc}
\toprule
\multirow{3}{*}{\textbf{SSL \& Surrogate}} & \multirow{3}{*}{\textbf{Attacker}} & \multicolumn{4}{c}{\textbf{Target model}} \\
\cmidrule(lr){3-6}
& & \multicolumn{2}{c}{Zhao19} & \multicolumn{2}{c}{Emo18} \\
\cmidrule(lr){3-4}\cmidrule(lr){5-6}
& & ASR (\%) $\uparrow$ & Speed (s) $\downarrow$ & ASR (\%) $\uparrow$ & Speed (s) $\downarrow$ \\
\midrule
\multirow{6}{*}{\makecell[c]{HuBERT\\ +\\ basemodel}}
  & PGD$_{0}$               & \textbf{99.42} & .0059 & \textbf{96.96} & .0065 \\
  & \textbf{SIGMA-PGD$_{0}$} & 95.67 & \textbf{.0050} & 92.47 & \textbf{.0040} \\
  \cmidrule(lr){2-6}
  & FW-$\ell_{1}$            & \textbf{99.42} & .0302 & \textbf{98.40} & .0323 \\
  & \textbf{SIGMA-FW-$\ell_{1}$} & 96.25 & \textbf{.0248} & 94.21 & \textbf{.0264} \\
  \cmidrule(lr){2-6}
  & Sparsefool        & \textbf{77.37} & .0461 & \textbf{98.70} & .0527 \\
  & \textbf{SIGMA-Sparsefool} & 80.98 & \textbf{.0401} & 98.12 & .0509 \\
\bottomrule
\end{tabular}
}
\end{table}

\begin{table}[htbp]
  \centering
  \small
  \setlength{\tabcolsep}{6pt}
   \caption {Black-box one-shot 0-query transfer on IEMOCAP. The surrogate is fixed to \textit{BaseModel} for each SSL. Adversarial examples are crafted once on the surrogate and applied to each target without any queries. Budgets are aligned within each SSL. Stars indicate a significant improvement over MI-FGSM on the same target by McNemar’s exact two-sided test ($^{*}p{<}.001$).}
  \label{tab:m3-all-ssl}
  \begin{tabular}{l l l l}
    \toprule
    \textbf{SSL} & \textbf{Target} & \textbf{Attacker} & \textbf{ASR (\%) $\uparrow$} \\
    \midrule
    \multirow{4}{*}{Emotion2Vec 
    }
      & \multirow{2}{*}{Zhao19}
        & MI-FGSM           & 69.71 \\
      & & \textbf{SIGMA-PGD$_{0}$}      & \textbf{75.34$^{*}$} \\
      \cmidrule(l){2-4}
      & \multirow{2}{*}{Emo18}
        & MI-FGSM           & 61.22 \\
      & & \textbf{SIGMA-PGD$_{0}$ }     & \textbf{67.07$^{*}$} \\
    \midrule
    \multirow{4}{*}{WavLM 
    }
      & \multirow{2}{*}{Zhao19}
        & MI-FGSM             & 61.46 \\
      & &\textbf{SIGMA-PGD$_{0}$}      & \textbf{69.36$^{*}$} \\
      \cmidrule(l){2-4}
      & \multirow{2}{*}{Emo18}
        & MI-FGSM            & 65.65 \\
      & & GI\_PGD$_{0}$  & \textbf{70.27$^{*}$} \\
    \midrule
    \multirow{4}{*}{HuBERT
    }
      & \multirow{2}{*}{Zhao19}
        & MI-FGSM            & \textbf{70.17} \\
      & & \textbf{SIGMA-PGD$_{0}$ }   & 62.82 \\
      \cmidrule(l){2-4}
      & \multirow{2}{*}{Emo18}
        & MI-FGSM          & \textbf{65.99} \\
      & & \textbf{SIGMA-PGD$_{0}$ }     & 58.03 \\
    \bottomrule
  \end{tabular}
\end{table}

We therefore next study the surrogate-to-target transfer SIGMA case, where a single mask is computed on a surrogate and reused on targets. Table \ref{tab:transfer} shows the result on IEMOCAP. It is important to clarify the experimental setup here to ensure a fair comparison. Standard sparse baseline methods (e.g., $PGD_0$, $FW-l_1$, Sparsefool) couple their sparse selection indices tightly with the target model's specific gradient landscape. Consequently, their support sets do not naturally transfer across models without recomputing the gradients or Top-K elements for each specific target. Therefore, the baseline results reported in this table are computed directly on each target model, representing a target-specific white-box upper bound.In contrast, our SIGMA-based approach fixes the perturbation support via a saliency mask M. We compute the mask once on a surrogate and reuse it on the target. Updates on the target are restricted to this transferred support.Under a matched budget, transferred SIGMA-PGD$_{0}$, SIGMA-FW-$\ell_{1}$ and SIGMA-Sparsefool achieve ASR close to their baselines. Some settings exceed the baselines; for example, HuBERT $\rightarrow$ Zhao19 with SIGMA-Sparsefool reaches 80.98\% (baseline 77.37\%). Average generation time over successful samples is typically comparable or lower. These results indicate that saliency-guided sparse support is compatible with multiple sparse attacks and remains effective under transfer. Reusing a single mask also amortises the total cost.

To better approximate realistic scenarios where the target model is inaccessible, we further evaluate SIGMA in transfer-based and black-box settings on the larger and more widely used dataset, that is, IEMOCAP. Table \ref{tab:m3-all-ssl} reports black-box one-shot 0-query transfer with the surrogate fixed to BaseModel. Compared with the strong baseline MI-FGSM, SIGMA-PGD$_0$ achieves a higher ASR than MI-FGSM when transferring to Zhao19 and Emo18 under Emotion2Vec and WavLM, with statistically significant gains. Under HuBERT, MI-FGSM remains stronger on both targets. These results suggest that saliency-guided sparse perturbations can improve transferability for some SSL families and highlight the value of SIGMA for zero-query transfer.

\begin{figure}[t]
  \includegraphics[width=\linewidth]{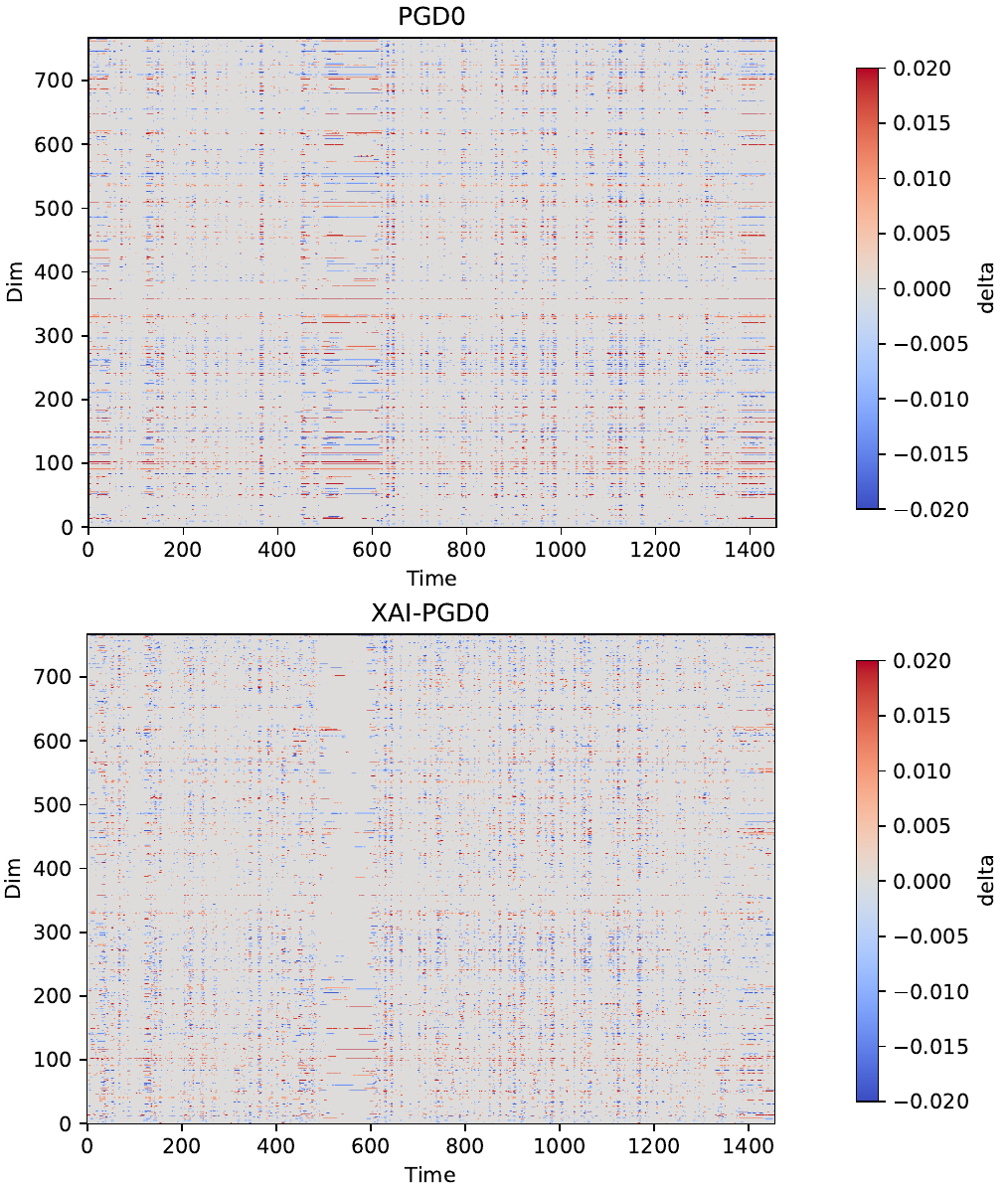}
  \caption{Perturbations on one test utterance (Emo2Vec+BaseModel) under the same budget ($\epsilon=0.02$, 10 steps, topk = 0.2). Top: baseline PGD$_{0}$; Bottom: SIGMA-PGD$_{0}$. }
  \label{fig:adv-two-panel}
\end{figure}

\begin{table*}[t]
\centering
\small
\caption{XAI ablation study on IEMOCAP. $\varepsilon=0.02$ }
\renewcommand{\arraystretch}{0.9}
\label{tab:sparse-PGD$_{0}$}
\begin{tabular}{lccccccc}
\toprule
\textbf{Method} & \textbf{Top-k} & \textbf{ASR (\%) $\uparrow$} & \textbf{$L_2$ Avg $\uparrow$} & \textbf{Sparsity $\uparrow$} & \textbf{Avg Time (s) $\downarrow$} &\textbf{Saliency Time (s) $\downarrow$}  &  \\
\midrule
\multirow{6}{*}{GI-PGD$_{0}$}
 & 0.02 &17.69  & .1872  & .0200 & .0066 &  10.03  &  \\
 & 0.05 & 36.16 & .2946 & .0499 & .0066 & 6.08 &  \\
 & 0.10 &  52.45  & .3991 & .0953 & .0066 & 6.20  &  \\
 & 0.20 & 64.87 & .4849 & .1562 &.0065  & 6.49  &  \\
 & 0.30 & 68.63 & .5125 & .1877 &.0065  & 6.66  &  \\
 & 0.40 &71.27  & .5233 & .2048 & .0065 & 6.80 &  \\
\midrule
\multirow{6}{*}{IG-PGD$_{0}$}
 & 0.02 & 15.05 & .1867 & .0200 &.0066  & 42.28 &  \\
 & 0.05 & 30.36 & .2928 & .0498 & .0067 & 41.78 &  \\
 & 0.10 &50.06& .3966 & .0951 & .0066 &  41.99 &  \\
 & 0.20 &64.24 & .4835 & .1560 & .0065 & 42.22 &  \\
 & 0.30 & 68.26 & .5117 &  .1876& .0065 & 42.12  &  \\
 & 0.40 & 70.89 & .5228 & .2047 & .0065 & 42.71    &  \\
\midrule
\multirow{6}{*}{LIME-PGD$_{0}$}
 & 0.02 &3.13  & .1389 & .0124& .0062 & 305.44 &  \\
 & 0.05 &11.04  &  .2260& .0336 & .0067 &303.94  &  \\
 & 0.10 & 34.50 &.3542  &.0797  & .0067 & 305.87  &  \\
 & 0.20 & 60.48 & .4733 &.1500  & .0065 & 303.58  &  \\
 & 0.30 &66.50  & .5084 & .1853 & .0065 &306.30   &  \\
 & 0.40 & 69.01 &.5215  & .2037 & .0065 &   306.07  &  \\
\bottomrule
\end{tabular}
\end{table*}

\begin{table}[t]
\centering
\caption{Ablation comparing SIGMA against a Random Sparse Mask on IEMOCAP ($\epsilon=0.02$).}
\label{tab:random_ablation}
\begin{tabular}{lcc}
\toprule
\textbf{Attack Method} & \textbf{Sparsity Rate ($k$)} & \textbf{ASR (\%)$\uparrow$} \\
\midrule
SIGMA-GI & 0.20 & \textbf{64.87} \\
Random Mask & 0.20 & 60.35 \\
\midrule
SIGMA-GI & 0.05 & \textbf{36.14 }\\
Random Mask & 0.05 & 21.46 \\
\bottomrule
\end{tabular}
\end{table}

\subsection{Ablation Study and XAI Analysis}

To validate the choice of saliency method and top-$k$ rate, we evaluate SIGMA-PGD$_{0}$ under a fixed $\varepsilon{=}0.02$ and a fixed number of iterations. We evaluate three post-hoc explainability methods (GI, IG, LIME) across several $k$ values (Table~\ref{tab:sparse-PGD$_{0}$}). The ASR increases monotonically with $k$. For GI-PGD$_{0}$, it rises from $17.69\%$ at $k{=}0.02$ to $71.27\%$ at $k{=}0.40$. IG and LIME show the same trend. At the same time, the $L_2$ average and sparsity also increase with $k$, which indicates higher perturbation energy and a higher perturbation sparsity.

For computational cost, the SIGMA-PGD$_{0}$ \textit{Avg Time} is nearly constant across $k$. The outer iteration count is fixed and each step applies element-wise updates with box clipping, so the mask rate has little effect on the cost of a forward or backward pass. In contrast, \textit{Saliency Time} differs markedly. GI needs about 6 to 10 seconds once. IG takes about 42 seconds due to path integration. LIME takes about 300 seconds because it relies on many local perturbations and surrogate fitting. This results in a massive total latency that restricts LIME's practical deployment for real-time single-sample attacks. GI therefore offers a favourable balance between accuracy and cost.

Taking both effectiveness and cost into account, $k{=}0.20$ is a balanced choice. Relative to $k{=}0.10$, the ASR improves notably (GI: from $52.45\%$ to $64.87\%$; IG: from $50.06\%$ to $64.24\%$; LIME: from $34.50\%$ to $60.48\%$). Further increases to $k{=}0.30$ or $k{=}0.40$ yield smaller gains and come with higher perturbation energy and sparsity.

In addition, to strictly validate that the attack efficacy stems from the saliency guidance rather than merely from the spatial sparsity constraint, we further conduct an ablation study comparing SIGMA (GI) against a Random Sparse Mask baseline on the Emotion2Vec model (Table \ref{tab:random_ablation}). At a moderate sparsity budget ($k=0.20$), the high-dimensional latent space provides sufficient degrees of freedom for the optimiser to find adversarial directions even within a random support. However, under a highly restrictive budget ($k=0.05$), the Random baseline suffers a severe collapse in ASR (dropping to 21.46\%), whereas SIGMA maintains a functional ASR of 36.14\%. This striking contrast confirms that randomly perturbing the latent space fails to intersect with the classifier's vulnerable directions under strict constraints.

Furthermore, we analyse explanation consistency. On \textsc{Emotion2Vec}, we evaluate white-box sparse PGD$_{0}$ using target-side attribution; results are reported in Table~\ref{tab:whitebox_drift_emotion2vec_noaopc} and \ref{tab:whitebox_drift_emotion2vec_noaopc_tess}. On IEMOCAP, compared with the gradient Top\,$k$ baseline, the GI/IG/LIME variants generally achieve higher TopK$\cap$ and Kendall’s $\tau_b$, and lower $\Delta\text{Sal}$, across BaseModel, Zhao19, and Emo18. 
The gains are moderate but statistically significant under paired tests
with Bonferroni correction. LIME delivers the largest consistency improvement, but it requires the highest saliency precomputation time; GI and IG offer a more balanced accuracy–cost trade-off. Online crafting time remains comparable to the baseline and ASR changes are small. We obesrve the similar trend on the TESS dataset. Overall, selecting the support with XAI and updating only within that support reduces attribution drift and improves explanation consistency. Furthermore, Figure \ref{fig:adv-two-panel} shows perturbations on one test utterance under the same budget. Both methods form band-like patches, but SIGMA-PGD$_{0}$ concentrates updates into fewer, cleaner bands and leaves wider zero regions, whereas PGD$_{0}$ spreads small updates more diffusely around the bands. This indicates higher energy concentration on salient elements for SIGMA-PGD$_{0}$ under the same bound.

\subsection{Acoustic Analysis of Saliency Masks}

To further examine whether the saliency masks capture affect-relevant information, we conduct a lightweight acoustic analysis. We focus on two fundamental prosodic cues: short-time energy (RMS) and fundamental frequency (F0). These are widely associated with emotional expression in speech . We process all utterances from the IEMOCAP dataset. We group them into high-arousal (e.g., angry, happy) and low-arousal (e.g., sad, neutral) categories. For each utterance, we extract frame-level RMS energy and F0 using standard signal processing tools. We align these with the corresponding SSL feature frames. We then compare the acoustic characteristics of frames selected by the SIGMA saliency mask (top 20\%) against those not selected. F0 statistics are computed over voiced frames only to avoid bias from unvoiced regions.

Table~\ref{tab:acoustic_analysis} summarises the results. Frames selected by SIGMA consistently exhibit higher energy and pitch levels than non-selected frames. This trend is highly pronounced in high-arousal emotions. Here, prosodic dynamics play a key role. For example, selected frames show an 81.7\% relative increase in RMS energy and a +10.9 Hz elevation in F0. In low-arousal emotions, the differences are milder but remain strictly consistent. These results suggest that saliency-guided masks do not select feature dimensions arbitrarily. Instead, they align with acoustically salient regions that correlate with emotional expression. We note that this analysis reflects statistical tendencies rather than causal relationships. Nevertheless, it provides strong supporting evidence. It shows that our framework focuses perturbations on regions consistent with established prosodic correlates of affect. This complements our quantitative evaluations. It provides an explainable perspective on why saliency-guided perturbations remain effective under strict sparsity constraints.

\begin{table}[htbp]
\centering
\caption{Acoustic characteristics of frames selected by SIGMA versus non-selected frames on the IEMOCAP dataset.}
\label{tab:acoustic_analysis}
\begin{tabular}{llcc}
\toprule
\textbf{Arousal Group} & \textbf{Feature} & \textbf{Selected} & \textbf{Non-Selected} \\
\midrule
\multirow{2}{*}{High-Arousal} 
& RMS Energy & 0.0482 $\uparrow$ & 0.0265 \\
& F0 (Hz)    & 224.4 $\uparrow$ & 213.5 \\
\midrule
\multirow{2}{*}{Low-Arousal}  
& RMS Energy & 0.0143 $\uparrow$ & 0.0085 \\
& F0 (Hz)    & 173.5 $\uparrow$ & 167.4 \\
\bottomrule
\end{tabular}
\end{table}

\section{Conclusion}
We introduced SIGMA, a modular framework that uses post-hoc attributions to restrict sparse adversarial updates to the most salient feature elements. Evaluations on two benchmark databases demonstrate its effectiveness across multiple speech encoders and attack methods. SIGMA perturbs substantially fewer features and reduces crafting time. It maintains competitive performance by navigating a controlled trade-off between attack success, sparsity, and explanation consistency. Furthermore, its saliency masks can be computed once on a surrogate model and efficiently reused. In zero-query settings, SIGMA improves transferability for Emotion2Vec and WavLM, indicating that saliency-guided masks capture structures that generalise across architectures.

Our evaluation targets the SSL feature space as a controlled analytical setting. This isolates the effects of attribution-guided perturbations without being confounded by waveform-level transformations, such as codecs. We do not assume real-world attackers can directly manipulate latent representations. Instead, SIGMA serves as a diagnostic framework for analysing model vulnerability and explanation behaviour in SER pipelines. While transfer remains stable under shared front-ends, representation shifts cause degradation across mismatched encoders. Future work will extend this framework to waveform-level attacks, cross-front-end generalisation, and human-centred evaluations. All code and models will be released to support reproducibility. All code, trained models, and attack scripts will be released to support reproducibility.

\bibliography{ref}

@inproceedings{shrikumar2017learning,
  author = {A. Shrikumar and P. Greenside and A. Kundaje},
  title = {Learning important features through propagating activation differences},
  booktitle={Proc.\ ICML},
  pages = {3145--3153},
  year = {2017}
}

@incollection{li2025explainable,
  title={Explainable AI for Healthcare},
  author={Li, Yupei and Sun, Qiyang and Akman, Alican and Schuller, Bj{\"o}rn W},
  booktitle={Handbook on Smart Health},
  pages={632--652},
  year={2025},
  publisher={SAGE Publications 1 Oliver's Yard, 55 City Road, London, EC1Y 1SP}
}

@inproceedings{sundararajan2017axiomatic,
  author = {M. Sundararajan and A. Taly and Q. Yan},
  title = {Axiomatic attribution for deep networks},
  booktitle={Proc.\ ICML},
  pages = {3319--3328},
  year = {2017}
}

@inproceedings{ribeiro2016should,
  author = {M. T. Ribeiro and S. Singh and C. Guestrin},
  title = {{``Why should I trust you?''} Explaining the predictions of any classifier},
  booktitle={Proc.\ KDD},
  pages = {1135--1144},
  year = {2016}
}

@inproceedings{ma2023emotion2vec,
    title = "emotion2vec: Self-Supervised Pre-Training for Speech Emotion Representation",
    author = "Ma, Ziyang  and
      Zheng, Zhisheng  and
      Ye, Jiaxin  and
      Li, Jinchao  and
      Gao, Zhifu  and
      Zhang, ShiLiang  and
      Chen, Xie",
    year = "2024",
    booktitle={Proc.\ ACL},
    pages = "15747--15760"
}

@article{busso2008iemocap,
  author = {C. Busso and M. Bulut and C.-C. Lee and A. Kazemzadeh and E. Mower and S. Kim and J. N. Chang and S. Lee and S. S. Narayanan},
  title = {{IEMOCAP}: Interactive emotional dyadic motion capture database},
  journal = {Language resources and evaluation},
  volume = {42},
  number = {4},
  pages = {335--359},
  year = {2008}
}

@inproceedings{tzirakis2018end,
  author = {P. Tzirakis and J. Zhang and B. W. Schuller},
  title = {End-to-end speech emotion recognition using deep neural networks},
  booktitle={Proc.\ ICASSP},
  pages = {5089--5093},
  year = {2018}
}

@article{yang2021superb,
  title={{SUPERB}: Speech processing {Universal PERformance} Benchmark},
  author={S.-w. Yang and P.-H. Chi and Y.-S. Chuang and C.-I. J. Lai and K. Lakhotia and Y.-T. Lin and A. T. Liu and J. Shi and X. Chang and G.-T. Lin and others},
  journal={arXiv preprint:2105.01051},
pages={1--6},
  year={2021}
}

@inproceedings{baevski2020wav2vec,
  author = {A. Baevski and Y. Zhou and A. Mohamed and M. Auli},
  title = {{wav2vec 2.0}: A framework for self-supervised learning of speech representations},
  booktitle={Proc.\ KDDNeurIPS},

  volume = {33},
  pages = {12449--12460},
  year = {2020}
}

@article{amiriparian2024exhubert,
  title={{ExHuBERT}: Enhancing {HuBERT} through block extension and fine-tuning on 37 emotion datasets},
  author={S. Amiriparian and F. Packa{\'n} and M. Gerczuk and B. W. Schuller},
  journal={arXiv preprint:2406.10275},
pages={1--5},
  year={2024}
}

@article{pepino2021emotion,
  title={Emotion recognition from speech using {wav2vec 2.0} embeddings},
  author={L. Pepino and P. Riera and L. Ferrer},
  journal={arXiv preprint:2104.03502},
pages={1--5},
  year={2021}
}

@article{nfissi2024unveiling,
  title={Unveiling hidden factors: Explainable {AI} for feature boosting in speech emotion recognition},
  author={A. Nfissi and W. Bouachir and N. Bouguila and B. Mishara},
  journal={arXiv preprint:2406.01624},
pages={1--36},
  year={2024}
}

@article{huang2023focus,
  author = {Q.-X. Huang and L.-K. Chiang and M.-Y. Chiu and H.-M. Sun},
  title = {Focus-shifting attack: An adversarial attack that retains saliency map information and manipulates model explanations},
  journal = {IEEE Transactions on Reliability},
  volume = {73},
  number = {2},
  pages = {808--819},
  year = {2023}
}

@article{raina2024muting,
  title={Muting {Whisper}: A universal acoustic adversarial attack on speech foundation models},
  author={V. Raina and R. Ma and C. McGhee and K. Knill and M. Gales},
  journal={arXiv preprint:2405.06134},
pages={1--18},
  year={2024}
}

@article{nieradzik2025reliable,
  author = {L. Nieradzik and H. Stephani and J. Keuper},
  title = {Reliable evaluation of attribution maps in {CNNs}: A perturbation-based approach},
  journal = {International Journal of Computer Vision},
  volume = {133},
  number = {5},
  pages = {2392--2409},
  year = {2025}
}

@article{haunschmid2020audiolime,
  title={{audiolime}: Listenable explanations using source separation},
  author={V. Haunschmid and E. Manilow and G. Widmer},
  journal={arXiv preprint:2008.00582},
pages={1--5},
  year={2020}
}

@article{akman2024audio,
  author = {A. Akman and B. W. Schuller},
  title = {Audio explainable artificial intelligence: A review},
  journal = {Intelligent Computing},
  volume = {2},
  pages = {0074},
  year = {2024}
}

@inproceedings{lundberg2017unified,
  author = {S. M. Lundberg and S.-I. Lee},
  title = {A unified approach to interpreting model predictions},
  booktitle={Proc.\ NeurIPS},
  volume = {30},
  year = {2017}
}

@inproceedings{selvaraju2017grad,
  author = {R. R. Selvaraju and M. Cogswell and A. Das and R. Vedantam and D. Parikh and D. Batra},
  title = {{Grad-CAM}: Visual explanations from deep networks via gradient-based localization},
  booktitle={Proc.\ ICCV},
  pages = {618--626},
  year = {2017}
}

@incollection{montavon2019layer,
  author = {G. Montavon and A. Binder and S. Lapuschkin and W. Samek and K.-R. M{\"u}ller},
  title = {Layer-wise relevance propagation: An overview},
  booktitle = {Explainable {AI}: Interpreting, Explaining and Visualizing Deep Learning},
  series = {Lecture Notes in Computer Science},
  volume = {11700},
  pages = {193--209},
  year = {2019},
  publisher = {Springer}
}

@article{sun2025explainable,
  author = {Q. Sun and A. Akman and B. W. Schuller},
  title = {Explainable artificial intelligence for medical applications: A review},
  journal = {ACM Transactions on Computing for Healthcare},
  volume = {6},
  number = {2},
  pages = {1--31},
  year = {2025}
}

@article{wang2021fine,
  title={A fine-tuned {wav2vec 2.0}/{HuBERT} benchmark for speech emotion recognition, speaker verification and spoken language understanding},
  author={Y. Wang and A. Boumadane and A. Heba},
  journal={arXiv preprint:2111.02735},
pages={1--7},
  year={2021}
}

@inproceedings{akman2025audio,
  author = {A. Akman and Q. Sun and B. W. Schuller},
  title = {Audio explanation synthesis with generative foundation models},
  booktitle={Proc.\ ICASSP},
  pages = {1--5},
  year = {2025}
}

@inproceedings{schuller2003hidden,
  author = {B. Schuller and G. Rigoll and M. Lang},
  title = {Hidden {Markov} model-based speech emotion recognition},
  booktitle={Proc.\ ICASSP},
  volume = {2},
  pages = {II--1},
  year = {2003}
}

@inproceedings{dong2018boosting,
  author = {Y. Dong and F. Liao and T. Pang and H. Su and J. Zhu and X. Hu and J. Li},
  title = {Boosting adversarial attacks with momentum},
  booktitle={Proc.\ CVPR},
  pages = {9185--9193},
  year = {2018}
}

@inproceedings{liu2020weighted,
  author = {X. Liu and K. Wan and Y. Ding and X. Zhang and Q. Zhu},
  title = {Weighted-sampling audio adversarial example attack},
  booktitle={Proc.\ AAAI},
  pages = {4908--4915},
  year = {2020}
}

@article{ghosh2022black,
  author = {A. Ghosh and S. S. Mullick and S. Datta and S. Das and A. K. Das and R. Mallipeddi},
  title = {A black-box adversarial attack strategy with adjustable sparsity and generalizability for deep image classifiers},
  journal = {Pattern Recognition},
  volume = {122},
  pages = {108279},
  year = {2022}
}

@article{dai2023saliency,
  author = {Z. Dai and S. Liu and Q. Li and K. Tang},
  title = {Saliency attack: Towards imperceptible black-box adversarial attack},
  journal = {ACM Transactions on Intelligent Systems and Technology.},
  volume = {14},
  number = {3},
  pages = {1--20},
  year = {2023}
}

@article{wiyatno2018maximal,
  title={Maximal {Jacobian}-based saliency map attack},
  author={R. Wiyatno and A. Xu},
  journal={arXiv preprint:1808.07945},
pages={1--5},
  year={2018}
}

@article{chang2024staa,
  author = {Y. Chang and Z. Ren and Z. Zhang and X. Jing and K. Qian and X. Shao and B. Hu and T. Schultz and B. W. Schuller},
  title = {{STAA}-net: A sparse and transferable adversarial attack for speech emotion recognition},
  journal = {IEEE Transactions on Affective Computing},
  year = {2024}
}

@inproceedings{liu2021audio,
  author = {X. Liu and X. Chen and M. Yin and Y. Wang and T. Hu and K. Ding},
  title = {Audio injection adversarial example attack},
   booktitle={Proc.\ ICML Workshop on Adversarial Machine Learning},
  year = {2021}
}

@article{jin2025boosting,
  title={Boosting the Transferability of Audio Adversarial Examples with Acoustic Representation Optimization},
  author={W. Jin and J. Su and H. Wang and Y. Ye and J. Hao},
  journal={arXiv preprint:2503.19591},
pages={1--16},
  year={2025}
}

@article{zhang2022adversarial,
  title={Adversarial attacks on {ASR} systems: An overview},
  author={X. Zhang and H. Tan and X. Huang and D. Zhang and K. Tang and Z. Gu},
  journal={arXiv preprint:2208.02250},
pages={1--8},
  year={2022}
}

@inproceedings{ren2020generating,
  author = {Z. Ren and A. Baird and J. Han and Z. Zhang and B. W. Schuller},
  title = {Generating and protecting against adversarial attacks for deep speech-based emotion recognition models},
  booktitle={Proc.\ ICASSP},

  pages = {7184--7188},
  year = {2020}
}

@article{lan2022adversarial,
  author = {J. Lan and R. Zhang and Z. Yan and J. Wang and Y. Chen and R. Hou},
  title = {Adversarial attacks and defenses in speaker recognition systems: A survey},
  journal = {Journal of Systems Architecture},
  volume = {127},
  pages = {102526},
  year = {2022}
}

@article{8294186,
  author = {N. Akhtar and A. Mian},
  title = {Threat of adversarial attacks on deep learning in computer vision: A survey},
  journal = {IEEE Access},
  volume = {6},
  pages = {14410--14430},
  year = {2018},
  doi = {10.1109/ACCESS.2018.2807385}
}

@article{ren2020adversarial,
  author = {K. Ren and T. Zheng and Z. Qin and X. Liu},
  title = {Adversarial attacks and defenses in deep learning},
  journal = {Engineering},
  volume = {6},
  number = {3},
  pages = {346--360},
  year = {2020}
}

@article{li2025gatedxlstm,
  title={{GatedxLSTM}: A multimodal affective computing approach for emotion recognition in conversations},
  author={Y. Li and Q. Sun and S. M. K. Murthy and E. Alturki and B. W. Schuller},
  journal={arXiv preprint:2503.20919},
pages={1--9},
  year={2025}
}

@article{sun2024towards,
  title={{Towards friendly {AI}: A comprehensive review and new perspectives on human-{AI} alignment}},
  author={Q. Sun and Y. Li and E. Alturki and S. M. K. Murthy and B. W. Schuller},
  journal={arXiv preprint:2412.15114},
pages={1--15},
  year={2024}
}

@article{hu2022acoustically,
  author = {J. Hu and Y. Huang and X. Hu and Y. Xu},
  title = {The acoustically emotion-aware conversational agent with speech emotion recognition and empathetic responses},
  journal = {IEEE Transactions on Affective Computing},
  volume = {14},
  number = {1},
  pages = {17--30},
  year = {2022}
}

@article{matsouliadis2025speech,
  author = {L. Matsouliadis and E. Siamtanidou and N. Vryzas and C. Dimoulas},
  title = {Speech emotion recognition and serious games: An entertaining approach for crowdsourcing annotated samples},
  journal = {Information},
  volume = {16},
  number = {3},
  pages = {238},
  year = {2025}
}

@article{li2021make,
  author = {H.-C. Li and T. Pan and M.-H. Lee and H.-W. Chiu},
  title = {Make patient consultation warmer: A clinical application for speech emotion recognition},
  journal = {Applied Sciences},
  volume = {11},
  number = {11},
  year = {2021}
}

@article{zhao2021self,
  author = {Z. Zhao and K. Wang and Z. Bao and Z. Zhang and N. Cummins and S. Sun and H. Wang and J. Tao and B. W. Schuller},
  title = {Self-attention transfer networks for speech emotion recognition},
  journal = {Virtual Reality \& Intelligent Hardware},
  volume = {3},
  number = {1},
  pages = {43--54},
  year = {2021}
}

@article{wang2023speech,
  title={Speech emotion diarization: Which emotion appears when?},
  author={Y. Wang and M. Ravanelli and A. Nfissi and A. Yacoubi},
  journal={arXiv preprint:2306.12991},
pages={1--7},
  year={2023}
}

@article{zhao2019speech,
  author = {J. Zhao and X. Mao and L. Chen},
  title = {Speech emotion recognition using deep 1{D} \& 2{D} {CNN LSTM} networks},
  journal = {Biomedical Signal Processing and Control},
  volume = {47},
  pages = {312--323},
  year = {2019}
}

@misc{pichora2020toronto,
  author = {M. K. Pichora-Fuller and K. Dupuis},
  title = {Toronto emotional speech set ({TESS})},
  year = {2020},
  howpublished = {Scholars Portal Dataverse},
  url = {https://doi.org/10.5683/SP2/E8H2MF}
}

@inproceedings{advexample2015,
  author = {I. J. Goodfellow and J. Shlens and C. Szegedy},
  title = {Explaining and harnessing adversarial examples},
  booktitle={Proc.\ ICLR},
  year = {2015}
}

@inproceedings{madry2018towards,
  author = {A. Madry and A. Makelov and L. Schmidt and D. Tsipras and A. Vladu},
  title = {Towards deep learning models resistant to adversarial attacks},
  booktitle={Proc.\ ICLR},
  year = {2018}
}

@article{8601309,
  author = {J. Su and D. V. Vargas and K. Sakurai},
  title = {One pixel attack for fooling deep neural networks},
  journal = {IEEE Transactions on Evolutionary Computation},
  volume = {23},
  number = {5},
  pages = {828--841},
  year = {2019}
}

@inproceedings{10889191,
  author = {J. Chen and Y. Dai and F. Huang},
  title = {{DiffAttack}: Imperceptible and transferable audio adversarial attack via diffusion model},
  booktitle = {Proc. IEEE Int. Conf. Acoust., Speech Signal Process. (ICASSP)},
  pages = {1--5},
  year = {2025}
}

@inproceedings{11023348,
  author = {X. Yuan and J. Zhang and F. Guo and K. Chen and X. Wang and S. Zhang and Y. Chen and D. Liu and P. Li and Z. Wang and R. Zhu},
  title = {{EvilHarmony}: Stealthy adversarial attacks against black-box speech recognition systems},
  booktitle = {Proc. IEEE Symp. Secur. Privacy (SP)},
  booktitle={Proc.\ S&P},
  pages = {4569--4587},
  year = {2025}
}

@inproceedings{NeekharaHPDMK19,
  author = {P. Neekhara and S. Hussain and P. Pandey and S. Dubnov and J. J. McAuley and F. Koushanfar},
  title = {Universal adversarial perturbations for speech recognition systems},
  booktitle={Proc.\ Interspeech},
  pages = {481--485},
  year = {2019}
}

@article{KIM2023109286,
  author = {H. Kim and J. Park and J. Lee},
  title = {Generating transferable adversarial examples for speech classification},
  journal = {Pattern Recognition},
  volume = {137},
  pages = {109286},
  year = {2023}
}

@inproceedings{9413467,
  author = {W. Zhang and S. Zhao and L. Liu and J. Li and X. Cheng and T. F. Zheng and X. Hu},
  title = {Attack on practical speaker verification system using universal adversarial perturbations},
  booktitle={Proc.\ ICASSP},
  pages = {2575--2579},
  year = {2021}
}

@article{goodfellow2020generative,
  author = {I. Goodfellow and J. Pouget-Abadie and M. Mirza and B. Xu and D. Warde-Farley and S. Ozair and A. Courville and Y. Bengio},
  title = {Generative adversarial networks},
  journal = {Communications of the ACM},
  volume = {63},
  number = {11},
  pages = {139--144},
  year = {2020}
}

@inproceedings{xie2021enabling,
  author = {Y. Xie and Z. Li and C. Shi and J. Liu and Y. Chen and B. Yuan},
  title = {Enabling fast and universal audio adversarial attack using generative model},
    booktitle={Proc.\ AAAI},
  pages = {14129--14137},
  year = {2021}
}

@inproceedings{pmlr-v97-guo19a,
  author = {C. Guo and J. Gardner and Y. You and A. G. Wilson and K. Weinberger},
  title = {Simple black-box adversarial attacks},
  booktitle={Proc.\ ICML},
  volume = {97},
  pages = {2484--2493},
  year = {2019}
}
\bibliographystyle{IEEEtran}

\end{document}